\documentstyle[12pt]{article}
\title{Fundamental constants in effective
theory}
\author{G.E. Volovik
\\Low Temperature Laboratory, Helsinki University of
Technology\\
P.O.Box 2200, FIN-02015 HUT, Finland\\
\\
 L.D. Landau Institute for
Theoretical Physics\\  Kosygin Str. 2, 117940 Moscow, Russia
}

\begin{document}
\maketitle

\begin{abstract}
{There is a discussion between  L. B. Okun,   G.
Veneziano and M. J. Duff, concerning the number of fundamental
dimensionful constants in physics \cite{Trialogue}. They
advocated correspondingly 3, 2 and 0 fundamental constants.
Here we consider this problem on example of the
effective relativistic quantum field theory, which emerges
in the low energy corner of quantum liquids and which reproduces
many features of our physics including Lorentz invariance, chiral
fermions,
gauge fields and dynamical gravity.  } \end{abstract}

\section{Introduction.}
 Experimental bounds on possible variation of fundamental constants
in nature
 are discussed in comprehensive review \cite{Uzan}. Since we live
well below
 the Planck energy scale, we canot judge which of them are
 really fundamental, i.e. cannot be derived from the underlying
 trans-Planckian physics. Here we discuss this problem using
the effective relativistic quantum field theory (RQFT) arising
as emergent phenomenon in quantum liquids \cite{PhysRepRev},
or in other condensed matter systems \cite{LaughlinPines}. Since the
trans-Planckian physics of a quantum liquid -- the microscopic atomic
physics
-- is well known at least in principle, this allows us to look at the
problem
of fundamental constants from the outside, i.e.  from the point of
view of an
external observer who does not belong to the low-energy world of the
effective RQFT.  This observer knows both the effective and
microscopic
physics. He or she knows the origin of the `constants' which enter the
effective theory, and thus can judge whether they are fundamental
within the
effective theory, and whether they remain fundamental at the more
fundamental
(i.e.  more microscopic) level.

The status of  fundamental constants in quantum liquids depends on
the energy
scale used: the low-energy scale where the effective theory is
applicable, or
the atomic scale of the "Theory of Everything". Within the effective
theory
the status of constants also depends on the observer who measures
them:  the
`inner' observer belonging to the world of the low-energy
quasiparticles,
or the external observer who belongs to the world of microphysics.
In Table
(\ref{FundamentalConstants}) we shall use the following criteria of
fundamentality. The first line shows whether (+) or not (-) the
constant is
fundamental from the point of view of an inner observer.  The
constant is
fundamental if it cannot be derived within the effective theory used
by the
inner onbserver.  The second line refers to the external
observer if he uses the effective theory only. For him the fundamental
constant is the phenomenological parameter. In complete equilibrium
of a
quantum liquid, and in the absence of external perturbations, this
phenomenological parameter is completely fixed and thus can be
considered as
a fundamental. Finally the third line
corresponds to Weinberg criterion \cite{WeinbergCriterium}: the
constant is not
fundamental if it can be derived from the parameters of microscopic
physics.

 In the Table we consider the status of following constants:
the Planck constant $\hbar$;  the speed of light $c$, i.e. the maximum
attainable velocity in the effective theory; and the Newton's
gravitational
constant $G$.  From 19 parameters of the Standard
Model we discuss only one -- the fine structure constant $\alpha$,
which has
a direct analog in superfluid $^3$He-A \cite{PhysRepRev}.

\begin{equation}
\matrix{
{\rm `constant'}  &\hbar  &G  & c  & \alpha\cr
{\rm effective~theory~-- ~inner~observer}  &+   & + &+ &-\cr
{\rm effective~theory~-- ~external~observer}  &+      &+ & - &-  \cr
 {\rm microphysics}          &+      &-           & -   &- \cr
  }
 \label{FundamentalConstants} \end{equation}

Let us now discuss the status of each of 4 constants.

\section{Planck constant $\hbar$}

Inspection of the RQFT emerging in quantum liquids demonstrates that
within
this scheme we have only one fundamental constant -- the Planck
constant
$\hbar$. It is fundamental both within the effective
RQFT at low energy and in the microscopic physics of liquids
(all signs in $\hbar$-column of Eq. (\ref{FundamentalConstants}) are
$+$).
Quantum mechanics is built in quantum liquids, which are
essentially quantum objects. It is the quantum zero point motion
of atoms which gives rise to the ground state of the $^3$He-A liquid,
where the low-energy fermionic and bosonic collective modes behave as
relativistic chiral
fermions, and gauge and gravitational fields. The limit $\hbar=0$
simply does not exist, since there is no vacuum at $\hbar=0$, and
thus no
effective theory.

In quantum liquids belonging to the same universality class of Fermi-
points
as $^3$He-A, all physical laws of our world are more or less
reproduced in
the low-energy corner, except for quantum mechanics. In principle, it
is not
excluded that in the more comprehensive systems,  quantum mechanics
will also
acquire the status of an emergent low-energy phenomenon
\cite{AdlerQM}, with
the minus sign in the third line of the $\hbar$-column. Since this
does not
happen in our quantum liquids, we have no solid basis for discussion
of the
possibility of varying $\hbar$.

\section{Speed of light}

The metric of the
effective `Minkowski' space-time, in which the "relativistic"
quasiparticles
 (fermions and bosons)  propagate along
the geodesic curves in anisotropic quantum liquids, has the following
local
structure in case of anisotropic quantum liquid:  $g_{\mu\nu}={\rm
diag}(-1,c_x^{-2},c_y^{-2},c_z^{-2})$.  The analog of
speed of light -- the maximum attainable speed for low-energy fermions
or/and bosons -- depends on the direction of their propagation. Since
for an
external observer there is no unique speed of light $c$, we have the
minus sign in the second and third lines of the $c$-column in Eq.
(\ref{FundamentalConstants}).

The question, which arises in case of the anisotropic
speed of light: What `speed of light' enters the Einstein relation
$M=mc^2$
between the rest energy and the mass of the object? More
generally: Which $c$ enters physical equations?

 The answer to this question is rather simple: the speed
of light
$c$ never enters explicitly any physically reasonable equation if
it is written in covariant and gauge invariant form. The speed
of light is hidden within the metric tensor, which is the relevant
dynamical
variable. For example, the above metric $g_{\mu\nu}={\rm
diag}(-1,c_x^{-2},c_y^{-2},c_z^{-2})$ enters the energy spectrum
of massive particle in the following way:
$E^2=M^2 +g^{ik}p_ip_k$, or  $g^{\mu\nu}p_\mu p_\nu + M^2=0$,
where
$M$ is the rest energy. In these equations there is no $c$.

Usually the mass is determined as the response of momentum
to velocity: $p_i=M_{ik}v^k$. Since the velocity of the particle
is
$v^i=dE/dp_i$, one obtains for the mass tensor: $M_{ik}=
Eg_{ik}$. If the linear reponse of the
momentum to velocity is considered one has
\begin{equation}
M_{ik}({\rm linear})\equiv M_{ik}({\bf v}=0) =Mg_{ik}~.
\label{MomentumLinearResponse}
\end{equation}
This equation contains the rest energy $M$, but does not
contain
$c$ and
$m$ explicitly. The same is valid for any equation which is
written in covariant form: they never contain $m$ and $c$
separately.

In the effective theory, the problem of the fundamenatlity of $c$ is
rather
peculiar. For an inner observer living in the liquid,
who uses the rods and clocks made of low-energy quasiparticles, the
"speed of
light" does not depend on the direction of propagation. Moreover it
does not
depend on the velocity of the inner observer with respect to the
liquid. This
happens because of the physical Lorentz contraction experienced by
his rods
made of quasiparticles; his clock made of quasiparticles experience
the
retardation of time. All this conspire to make the inner observer
to believe that  the speed of light is fundamental. Such
low-energy observer  can savely divide the rest energy $M$  by his
$c^2$. He will obtain what he thinks is the mass of the object,
and this belief will be shared by all the low-energy inner
observers. In this sense the speed of light is fundamental and the
first
line of Table (\ref{FundamentalConstants}) contains the plus sign.

However, for the external observer who lives outside the
quantum liquid and uses the
rigid rods and laboratory clocks in his experiments,
the speed of massless quasiparticles is anisotropic even in the
low-energy limit, i.e. in the range of applicability of the effective
theory.
He or she finds that, say, in $^3$He-A the "speed of light" varies
from 3
cm/s to 60 m/s depending on the direction. This means that for
external
observer the speed of light is not fundamental both in the
microscopic and
effective theories (the minus sign both in the second and third lines
of
(\ref{FundamentalConstants})).

\section{Newton's constant}

 In some (very special) cases
\cite{PhysRepRev} the action for the effective gravitational field
in quantum liquids is similar to the Einstein action
\begin{equation}
S_{\rm Einstein}= - \int d^4x{\sqrt{-g}\over
16\pi G}{\cal R}
\label{EinsteinAction}
\end{equation}
As in the case of Sakharov's induced gravity \cite{Sakharov1967}, the
analog
of the gravitational constant $G$ in quantum liquids is determined by
the
Planck energy cut-off $E_{\rm Planck}$ (played by the amplitude of the
superfluid gap $\Delta_0$) and by the number of the chiral fermionic
species:
$G^{-1}\sim N_F E_{\rm Planck}^2/\hbar$.  Since $G$ is determined by
the
microscopic physics, it is not fundamental on the microscopic level --
 minus
sign in third line.

In quantum liquids the effective $G$ depends  on
temperature \cite{PhysRepRev},
$G^{-1}(T)=G^{-1}(0)+ \gamma N_F T^2/\hbar$,
where $\gamma$ is the dimensionless factor of
order unity.
Since the temperature correction does not contain the microscopic
parameters and thus in principle can be obtained within the effective
theory, $G$ is not fundamental even on the effective level.
But its vacuum value $G(T=0)$ is fundamental for an inner
observer.

Moreover, the phenomenological parameter $G(T=0)$ of the effective
theory as
viewed by the external observer has a definite value for the quantum
liquid isolated from the environment. That is why it can be
considered as the
fundamental constant of the effective theory, whence the plus sign in
the
second line of $G$-column.

\section{Fine structure constant}

In logarithmic approximation, the action for effective electrodynamics
emerging in $^3$He-A (and in other systems of the Fermi-point
universality
class) is \cite{PhysRepRev}:  \begin{equation} S_{\rm em} =
{1\over\hbar}\int
dt d^3x {\sqrt{-g}\over 16\pi \alpha}F^{\mu\nu}F_{\mu\nu}~.
 \label{Action3He} \end{equation} An analog of the fine structure
constant
$\alpha$ is  not fundamental both on microscopic and effective
levels, since
it logarithmically depends both on ultraviolet and on infra-red cut-
off
parameters. If the largest infra-red cut-off is supplied by
temperature, one
has $\alpha^{-1}\sim N_F \ln (E_{\rm Planck}/T)$, where $N_F$ is
again the
number of chiral fermionic species, and the ultraviolet cut-off is
determined
by the same Planck scale $E_{\rm Planck}=\Delta_0$ which enters the
effective
Newton's constant \cite{PhysRepRev}.  It is the same effective
electrodynamics as was discussed by Zeldovich \cite{Zeldovich1967b}.
Dependence on $E_{\rm Planck}$ shows that $\alpha$ is not fundamental
on the
microscopic level; the dependence on the low-energy scale
demonstrates that
it is not fundamental within the effective theory too. In contrast to
Newton's constant, in the limit of $T=0$ and zero mass of all the
Standard
Model fermions the constant $\alpha$ vanishes. That is why the
$\alpha$-column contains only the minus signs  in Eq.
(\ref{FundamentalConstants}).

Equation \ref{Action3He} does not contain speeds of light
$c_x$, $c_y$ and $c_z$
explicilty: these functions are hidden in the metric field. The
anisotropic speed of light is the crucial argument that
the decomposition of the dimensionless running coupling $\alpha$
into dimensionful $e^2$, $\hbar$ and  $c$ is meaningless:
$e^2$ and $c$ do not enter explicitly any covariant
and gauge invariant equations. For example, the energy levels of
the electron  in the hydrogen atom is determined by $\alpha$
and by the rest energy $M_e$ of the electron \cite{WeinbergBook}:
\begin{equation}
 E_{n,j} =
M_e\left[1- {\alpha^2\over 2n^2} - {\alpha^4\over
2n^4}\left({n\over j+1/2}-{3\over 4}\right)+...\right]
\end{equation}
Let us recall that in contrast to the quantity $e$, the physical
electric (or
other) charge of elementary particles in Standard Model and its
analog in
$^3$He-A are dimensionless (integer or fractional) quantities
determined by
geometry.  They are determined either by gauge groups in fundamental
qauge theories or by momentum-space topology in quantum vacua of the
Fermi-point universality class.

The only fundamental constant which enters explicitly the
electromagnetic action is $\hbar$.

\section{What constants are really fundamental?}

Thus, in our example of the relativistic quantum field theory
arising in quantum liquids, the only fundamental constant is
$\hbar$. Since it is fundamental it is possible to use it as a
conversion factor of energy to frequency as is stated by Duff
in \cite{Trialogue} (of course, with the following reservations:
(i) energy is additive while frequency is not; (ii) as distinct
from frequency the magnitude of energy depends on the choice of
zero energy level; (iii) the frequency can be bounded from above
due to, say, discreteness of time, while the energy is not
bounded: it can be the sum of energies of many low-energy particles;
etc.).

However, it is not completely excluded that $\hbar$ can arise
as a result of a more fundamental theory above the Planck scale
which gives rise  to the quantum mechanics in the low-energy
scale \cite{AdlerQM}.

In such case the only fundamental constants which remain
in the effective RQFT are the dimensionless charges and
dimensionless topological quantum numbers
\cite{ThoulessBook,PhysRepRev}
which enter the topological terms in effective action. Such terms are
naturally dimensionless, since they are determined either by gauge
groups
or by the topological invariants in real and in momentum spaces. In
3+1
systems, such invariants describe the chiral anomaly
\cite{Adler1969,BellJackiw1969}, and other anomalies. In 2+1 systems,
these
topological quantum numbers lead to quantization of physical
parameters such
as Hall and spin Hall conductivities, which being written in the
covariant
form are integer of rational numbers.

\vfill\eject

\end{document}